**Experimental and numerical analysis of a full-scale timber-concrete-composite beam from simply supported to frame-connected.**

Authors: Dolores Otero-Chans, Félix Suárez-Riestra, Emilio Martín-Gutiérrez, Javier Estévez-Cimadevila
Affiliation: Universidade da Coruña, GEA-GEM, ETSAC, Centro de Innovación Tecnolóxica en Edificación e Enxeñaría Civil (CITEEC). Campus de Elviña, 15071. A Coruña, Spain/España.

Summary

Numerous TCC (timber-concrete composite) systems have been proposed and studied in recent years. Most of the studies have focused on the behaviour of the timber-concrete shear connection and the flexural behaviour of floor elements. This paper presents the results of an experimental and numerical study comparing the behaviour of a simply support beam and a beam connected in continuity to mixed TCC supports (a frame-connected beam). Full-scale tests were carried out with 370 mm wide beams and 5.9 m span. Experimental results have shown that the deflection in the frame-connected beam was less than 50% of those observed in the simply supported beam. A three-dimensional finite element model has been validated and calibrated from the test results. This model has been used to evaluate the moment redistribution in the frame-connected beam which, in the case studied, allows the maximum stresses in the timber members to be reduced by up to 49%. Based on the numerical and experimental results, a dimensioning criterion is proposed for the frame-connected beams.

1. Introduction

Climate change is a global concern. In the so-called Paris Agreement, signed at the United Nations Climate Change Conference in 2015, the European Union made a commitment to move towards carbon neutrality in the second half of the 21st century. The building sector plays an important role in carbon emissions, so it is vital to work on solutions that enhance the efficient use of energy and resources in the construction and refurbishment of buildings [1].

Timber-concrete composite structural systems, commonly known as Timber-Concrete Composites (TCC), offer great advantages in this area. The synergy among timber and concrete, especially in bending systems, makes it possible to take advantage of the high strength of concrete in compressed areas and of timber in tensioned areas, reducing the volume of material required when compared to timber-only or concrete-only systems. TCC systems are stiffer, have better vibration performance and sound insulation than timber-only systems and, at the same time, are lighter and more sustainable systems than concrete-only solutions [2]. It is well known that timber as a building material is renewable, allows shortening construction times and reducing transport and waste, fixes $CO_2$ in the atmosphere during the life cycle of buildings and facilitates the industrialisation of technological products using little energy and limiting the use of fossil derivatives [3,4]. Regarding concrete, the thickness of the layers used in TCC systems is very small, which allows the advantages described above to be obtained while minimising the impact of the solution.

The effectiveness of TCC systems lies in the design of the timber-concrete connection, and its ability to transmit the shear stresses by limiting the relative slip between the two materials. Numerous studies and designs have been carried out on this shear connection, which can generally be grouped into three main types: adhesive connections [5-7], mechanical connections using lag bolts, bolts, angle brackets or steel tubes, etc. [8-11] and joints made by notches or indentations in the timber [12,13]. Generally, joints made with notches or adhesives have the highest stiffness values [2]. Our research group has worked on the characterisation of a shear connection system using drilled holes through timber ribs or webs that are filled by fresh concrete, generating a timber-concrete interaction that allows a solution with high stiffness values in service and ductile failure [14]. This solution is also more sustainable than those requiring additional adhesives or metal fittings, since it is solved by using only timber elements and the concrete of the top layer. Compared to the usual notched systems, it has the advantages of not requiring additional elements to ensure the transmission of possible tension effects between the timber and the concrete while offering a ductile failure.

In addition to the numerous experimental studies carried out on the behaviour of different types of shear connections, bending floor elements have also been studied experimentally, either in rib format [5] or in slab format [6]. Systems designed with a timber bottom slab offer the advantage that the timber is used also as a permanent formwork for the concrete top slab. These systems are usually made with cross-laminated timber [10,12,13], LVL [6] or glued laminated timber (glulam) members [7,11]. Our research group experimentally studied the behaviour of floor systems consisting of plywood ribs or webs (in which the holes are drilled for the shear connection to the concrete), a thin glulam bottom flange and a thin fibre-reinforced concrete top slab. The space between the ribs is filled with lightweight blocks (Fig. 1), which were made with cardboard in our previous studies, but can be made with any lightening material. The assembly have proved to provide a lightweight solution, achieving high inertia with a reduced volume of materials and, in addition, allowing high loads to be supported with high stiffness in simply supported systems [15]. The continuity of the concrete in the connection of all the component parts of the system (floors, beams and supports) would increase the overall stiffness of the structure compared to other only-timber solutions, or to TCC systems that work as simply supported.

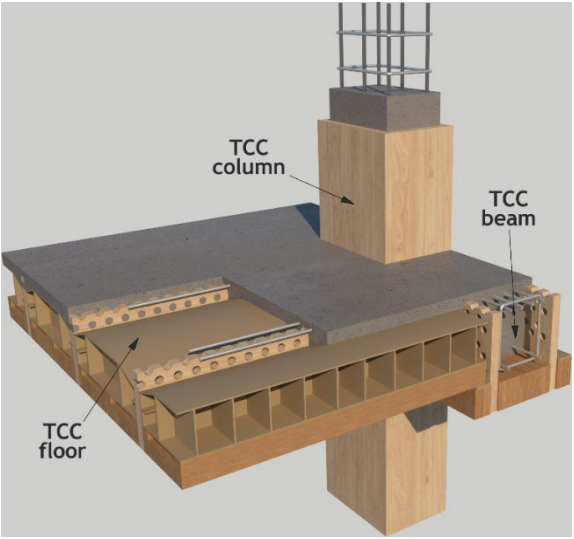

*Fig. 1 Global timber-concrete system. Includes TCC supports, beams and floors.*

The vast majority of existing studies have so far focused on the experimental analysis of simply supported systems [5,6,10], which implies a waste of one of the advantages offered by the use of concrete top slabs: the possibility of designing continuous floor slab solutions over successive spans. The use of continuous solutions reduces bending deflections, which in turn reduces the required floor height and thus the volume of material required. This continuity can also be applied to the floor-to-beam and beam-to-column connections, which would result in structural systems that are stiffer also against lateral actions and with better buckling behaviour due to the restriction of rotation in the joints. Fragiacomo [16] carried out full-scale tests of a floor with a conventional solution of timber joist connected with lag bolts to a concrete top slab. He made full-scale tests in an experimental building, comparing end-pinned joints and connections with restricted rotation through the joist-to-wall joint, finding for the latter case a reduced degree of stiffness of the connection. Kuhlmann & Schänzlin [17] carried out tests to analyse the continuity behaviour of TCC floors, but using steel beams instead of TCC beams. Continuity in bending elements also involves consideration of possible concrete cracking in negative bending areas (hogging bending). It has been found that, in connections made with screws, concrete cracking is associated with a lower efficiency of the connections in negative bending areas [18], although the incidence of this effect varies depending on the type of connector used. Beams made using glued-in steel meshes as shear connectors have shown clearly lower stiffness in the case of bending that generates tension in the concrete [19].

TCC floor systems have also been studied in relation to their behaviour as a rigid diaphragm associated with the design of buildings in seismic zones. Smith et al [20], studied the use of post-tensioned timber portal frames as an alternative for designing structures in seismic areas. They proposed the rigid diaphragms formed by the TCC floors to increase the stiffness of the system, but did not include concrete in the design of the frames or nodes. In the Timber Tower Research Project at the University of Oregon [21], an experimental study was carried out on continuous floor elements with TCC cross-sections, using continuous cross-laminated timber boards as a base; an analytical model for the design of portal frame structures considering all-timber columns in the portal frame design was carried out. Generally, the assessment of the use of concrete or TCC supports in the stiffness of the system is a neglected aspect in these studies.

Zhang et al [22] conducted an experimental study to evaluate the seismic behaviour of beam-to-column joints. They used glulam columns and beams. The beams were connected to a reinforced concrete slab by means of self-tapping screws. They evaluated six different beam-to-column joints, all made of steel plates. They concluded that the design of the plates has a great influence on the behaviour of the joints, but concrete is not used in the supports or in the design of the joint. Ribeiro et al [23] carried out an experimental study in which cast-in-place concrete joints are used to connect timber beams and columns to which steel rebars have been previously glued. They compare these connections with conventional bolted beam-to-column connections. The joints formed with glued bars and then poured were found to be much stiffer and stronger than the bolted joints. The required spacing between bolts and the cross-sections of the structural elements also limit the number of bolts that can be used and, consequently, the strength and stiffness of this type of joints. The solution proposed by Ribeiro et al offers a promising solution, even though multiplies the number and complexity of operations to be carried out and does not take advantage of the possibility of extending the continuity of the concrete to the connection between floors and beams.

Our research team proposed a structural system in which all the elements (supports, beams and floors) are designed using the TCC solution based on the shear connection made by means of holes drilled in plywood boards [24]. This would make it possible to take advantage of the continuity effect of concrete and to design joints with high moment carrying capacity in all connections between structural elements (Fig. 1). In this way, due to the increased bending stiffness and lateral stability of the system, the required cross-sections for the structural elements could be reduced compared to simply supported systems and additional bracing systems to resist lateral actions could be dispensed with. In previous articles, the flexural behaviour of simply supported floor elements with spans between 6.2 and 8.6 m [15] and cantilever elements [25] was studied. This paper shows the results from an experimental full-scale and numerical analysis of a simply supported beam and a beam connected to end supports forming a frame (frame-connected beam), carried out with the perforated board shear connection system. The increase in stiffness of the solution with semi-rigid beam-to-column connections with respect to the simply supported beam and what this may represent in the overall design of the structural system of buildings is shown.

## 2. Materials and methods

### 2.1. Specimens detail

A simply supported beam (SS) and a beam connected to upper and lower support portions forming a frame, called frame-connected beam (FC), were constructed to be tested. The total length of the simply supported beam specimen was 6.0 metres, to be tested with a 5.9-metre span between supports. The frame was designed with a 5.9-metre span between support axes for comparison with the simply supported beam.

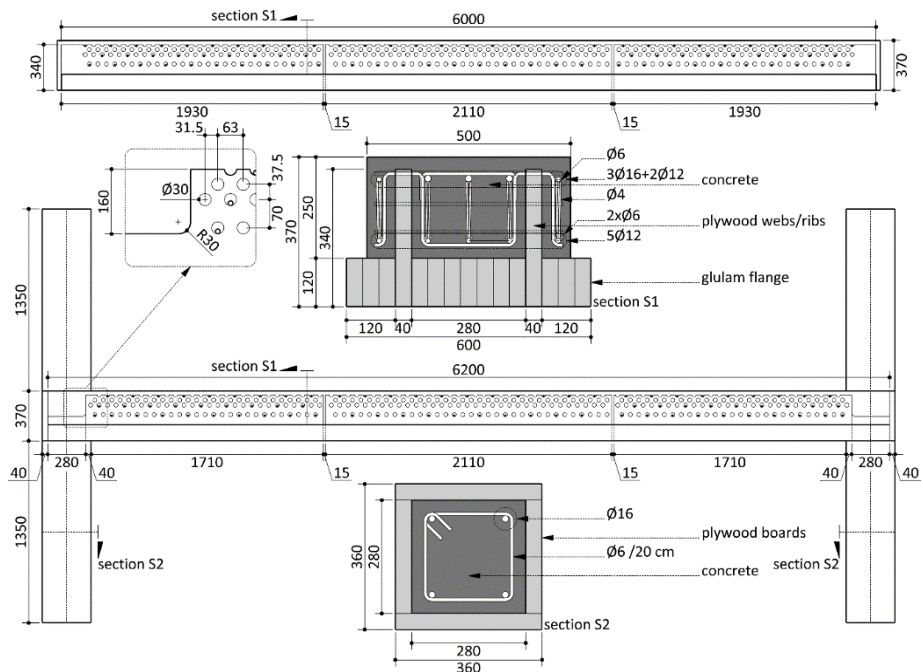

*Fig.2 Dimensions of the beams and the supports.*

The beams were formed with a bottom flange of glued laminated timber 120 mm thick and 600 mm wide. Two 40 mm thick birch plywood boards were sandwiched between the bottom flange, spaced 320 mm apart. The boards had a total height of 340 mm and were drilled with 3.5 lines of 30 mm diameter holes in staggered rows to allow connection to the fresh concrete, according to the pattern shown in the Fig. 2. Longitudinally, the boards were made up of three pieces that are simply attached to the ends, without any connecting element between them. The central piece was 2.13 metres long and the end pieces were 1.93 metres long. This system of discontinuous web makes it possible to easily apply a precamber to the beams (and to the floors, if necessary) before the concrete is poured, since the stiffness of the cross-section in this phase is limited to that of the lower flange of glulam at the points of discontinuity of the boards. The concrete section, with an external dimension of 250x500 mm, surrounded the boards and incorporated the reinforcement bars. This concrete section provided a 7 cm lateral overlay to the boards which housed a layer of reinforcement bars. The width of the concrete section (500 mm) was less than that of the timber section (600 mm) to simulate the support of the floor elements perpendicular to the beams that would occur in a real structure. This section, according to the system described in Fig. 1, would correspond to floor slabs with a total height of 25 cm, suitable for spans of 6 m in both residential and public use [15].

In the case of the frame-connected specimen, the beam was connected at each end to two 1350 mm long supports. This support length represents half the usual height between floors and corresponds to the approximate location of the null moment point under the usual hypothesis of gravity actions. The supports consisted of an outer box made of 40 mm thick birch plywood boards, in which circular notches had been made in a pattern similar to the holes drilled in the beams (Fig. 3). These boxes are designed to act as support and prompt for the beams during the assembly process of the structure and, after incorporating rebars, the fresh concrete is poured into them, generating the continuous connection of the system. In the case of a real building, the reinforced concrete also forms a continuous connection between the beams and the different floor elements (Fig. 1).

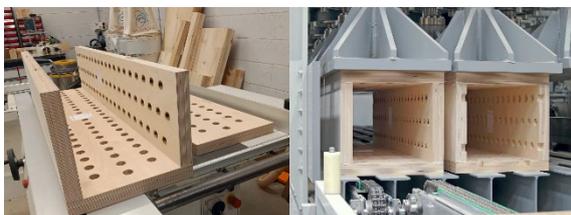

**Fig. 3** *Manufacturing process of the plywood board boxes that make up the outer layer of the TCC supports.*

The supports were reinforced with 4ϕ16 mm longitudinal bars and 6 mm diameter transversal rebars every 20 cm. The beams were reinforced with 3ϕ16 mm at the top and 3ϕ12 mm at the bottom along their entire length in the area between the boards. Double lattice girders, usually used to form floor slabs in reinforced concrete structures, with diagonals of ϕ4 mm and top and bottom mounting bars of ϕ6 mm, were used as shear reinforcement. The upper mounting bar is unique for both trusses. Between the boards, 3 double trusses were arranged to coincide with the 3 longitudinal reinforcement bars. On each overlaying area of the boards, 1ϕ12 mm at the top and 1ϕ12 mm at the bottom were arranged longitudinally, with a double shear reinforcement lattice girder equal to those used in the area between the boards. In addition, to resist the negative bending (hogging) moments of the beam, two additional reinforcement bars with a diameter of 16 mm were placed at each end close to the supports, with a length of 30+130 mm. Fig. 4 shows the pouring process and the reinforcement arrangement. This negative bending reinforcement was also placed in the simply supported beam, despite the lack of negative moments, in order to establish subsequent comparisons with the frame-connected beam.

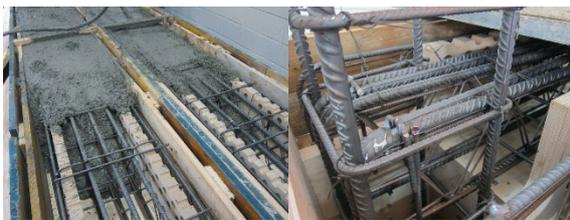

*Fig. 4 Concrete pouring in simply supported and frame-connected beams.*

Transversally, 8 mm diameter rebars were placed across the holes drilled in the boards, in order to increase the ductility of the shear connection, as observed in previous characterisation tests of these connection system [14]. The distribution of these bars, whose density increases in the extreme thirds of the beams, can be seen in Fig. 2.

The beams were poured with a 12 mm precamber (a 500 length-to-precamber ratio) in order to counteract possible effects derived from concrete shrinkage and deflection produced by the self-weight. At the time of assembly of the elements for the load tests, it was found that the deflection of the beams was in both cases less than 1 mm under self-weight effects. In relation to concrete shrinkage, its effect would be different in the case of multi-span continuous beams, due to the positive effect of the shortening that would occur in the areas of negative moments. This is a subject of great interest and deserves further detailed study.

## 2.2. Materials

The glulam bottom flange was made of Picea Abies of strength class GL24h, according to standard EN14080 [26]. Characteristic strength and stiffness for this class are: bending strength 24.0 N/mm$^2$; tension parallel to the fibre strength 19.2 N/mm$^2$; shear strength 3.5 N/mm$^2$; mean MOE 11600 N/mm$^2$.

Birch plywood was used for the webs. According to manufacturer specifications, average characteristic mechanical parameters parallel to the fibre are: compression strength 26.5 N/mm$^2$; tension strength 38.3 N/mm$^2$; panel shear 9.5 N/mm$^2$; mean MOE in bending 8925 N/mm$^2$ and mean MOR in shear 650 N/mm$^2$.

Reinforcement steel bars B500S quality were used with a yield strength $f_y \geq 500$ N/mm$^2$ and an ultimate tensile strength $f_u \geq 550$ N/mm$^2$ according to UNE 36068 [27]. Lattice girder as UNE 36739 [28] was used as shear reinforcement. The upper slab was made using fibre-reinforced concrete, with 360 kg/m$^3$ of CEM II/A-M (V-L) 42.5 R cement [29] and a proportion of 4% Sikafiber Force M-48 (Sika®) polyolefin macrofibres of 48 mm long and 465 N/mm$^2$ tensile strength, according to EN 14889-2 [30]. An average strength at 28 days of 38.1 N/mm$^2$ was obtained in compression tests carried out according to EN 12390-3 [31]. According to Eurocode 2 [32], modulus of elasticity has been determined as 34796 N/mm$^2$.

## 2.3. Test methods

Specimens were tested in bending in a four-point bending test configuration. The tests were carried out in the PEMADE laboratory of the University of Santiago de Compostela. A PB2-F/600 Microtest hydraulic machine was used. In the case of the frame, three load cells were used. Two of them, with a capacity of 200 kN, were used to apply load on the supports. The third, with a capacity of 600 kN, was used to apply the load on the beams. The loading was carried out by displacement control. The loading speed was adjusted so as not to exceed 2-3 minutes in duration for the non-destructive tests and 5-7 minutes in duration for the failure tests. The simply supported beam specimen was arranged over steel members with a span of 5.9 metres. A steel profile was used to distribute the load to two application points arranged at thirds of the span. Fig. 5 shows the arrangement of LVDT displacement sensors, three to measure the vertical displacement at the bottom edge of the beam and two the relative horizontal displacement at the ends of the beam, coinciding with timber-concrete interface. An LVDT was placed at the midpoint of the span, which corresponds to the maximum deflection measurement, and two other LVDTs were placed at a distance of 1.8 m that allow the evaluation of the deformation in the pure bending area (without shear). Five non-destructive tests were carried out, according to the maximum load values given in Table 1. The table also shows the equivalent distributed load values in terms of deformation, in order to establish comparisons with usual load situations represented by uniformly distributed loads over the beam. After reaching the maximum load for each test, the beam was unloaded and the remaining deflection value at that moment was recorded. Immediately the next test was started. Obviously, the remaining deflection would be lower if a longer period of time was set between the different tests, so it cannot be considered as representative. A first load of reduced value is carried out in test SS1, which can be considered as a preload. Deformations somewhat higher than in the subsequent tests are recorded in SS1 test, which could be due to a rearrangement of the different components of the system, and which are not considered significant as they would occur during the construction phase in a real building. In the non-destructive tests SS2 to SS5 the deformations suffered by the beam were recorded, which in the SS6 test was taken to rupture. In the failure test, the only the LVDT2 was kept in the midspan until the limit load corresponding to the SS5 test was reached, after which it was also removed to prevent it from being damaged by failure of the specimen.

In the case of the frame specimen, connecting fittings were provided at the base of the supports to prevent displacement but allow rotation (Fig. 5). To fix the head of the supports, a series of triangulated steel profiles were provided linked to the sides of the load cell support. This top connection system could not guarantee the total absence of displacement at this point, so LVDT1 was used to record the lateral displacement of the support head. For the frame test, a total of 8 LVDTs were provided, distributed as shown in Fig. 5. The six tests performed for the simply supported beam were repeated, using the same maximum load values given in Table 1. Test FC1 was performed with no load on the supports. Test 2 was carried out firstly without load on the supports (test FC2a), then applying a load of 100.0 kN on each support (test FC2b) and, finally, applying a load of 200.0 kN on each support to simulate a real working situation (test FC2c) and to see the influence that the axial force on the supports could have on the beam-support node rotational restraint. The successive tests were carried out maintaining this load on the supports. In the FC6 test, the aim was to take the frame to failure, so no LVDT records were made, although a failure in the laboratory slab on which the load cells were anchored forced the test to be stopped without reaching failure. After this maximum load test, two new loading phases were carried out, tests FC7a and FC7b, up to a maximum of 173.12 kN in order to evaluate the remaining capacity of the portal frame after reaching load values close to the limit. In the case of test FC7a no load was applied on the columns and in the case of test FC7b a load of 200.0 kN was maintained on each support. In tests FC7a and FC7b only the mid-point displacement (LVDT 2) was recorded.

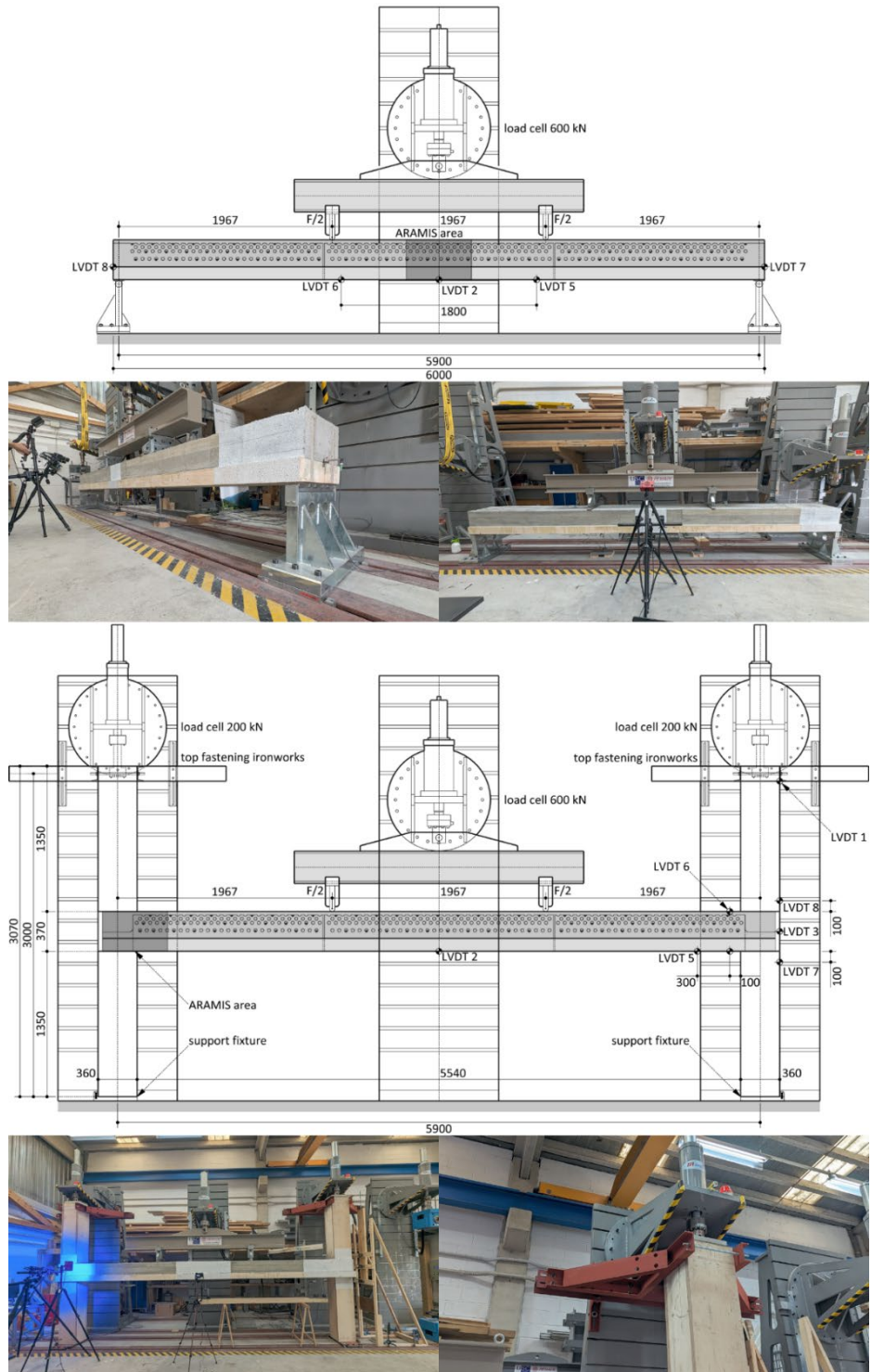

*Fig. 5* Test set-up for the simply supported (SS) beam and the frame-connected (FC) beam.
*Table 1*
*Value of maximum load applied in the different tests and equivalence in terms of deformation with uniformly distributed loads over the simply supported beam.*

| Test | | Load over the columns in FC test [kN] | Point loads (F/2) [kN] | Total load (F) [kN] | Equivalent distributed load for deformation effects [kN/m] |
|---|---|---|---|---|---|
| Simply supported beam | Frame-connected beam | | | | |
| SS1 | FC1 | 0.00 | 21.64 | 43.28 | 10.0 kN/m |
| SS2 | FC2a | 0.00 | 43.28 | 86.56 | 20.0 kN/m |
|  | FC2b | 100.00 | | | |
|     | FC2c | 200.00 |        |        |            |
| SS3 | FC3  | 200.00 | 64.92  | 129.84 | 30.0 kN/m  |
| SS4 | FC4  | 200.00 | 86.56  | 173.12 | 40.0 kN/m  |
| SS5 | FC5  | 200.00 | 108.20 | 216.40 | 50.0 kN/m  |
| SS6 | FC6  | 200.00 | Load until failure |        |   |
| -   | FC7a | 0.00   | 86.56  | 173.12 | -          |
|     | FC7a | 200.00 |        |        |            |

## 2.4. Vibrational behaviour

An experimental measurement of the dynamic response of the beams was carried out. For this purpose, the beams were excited by an impulse generated by an instrumented hammer and the response was measured with a digital accelerometer with an ADXL345 sensor, a measurement range of ±2G and an output data rate of 400 Hz. Measurements were taken prior to the start of the tests, one measurement in the simply supported beam and two measurements in the frame-connected beam, with no load on the columns and with a load of 200.0 kN on each support.

## 2.5. Digital Image Correlation

A Digital Image Correlation (DIC) system was used, using the ARAMIS measurement system. The data collection areas were located at the midspan in the case of the simply supported beam, and at the beam-column joint in the case of the frame-connected beam (Fig. 5).

In the case of the frame-connected beam, the behaviour of 7 lines/sections corresponding to the support axis (LA), 4 lines parallel to it at different distances (L1 to L4) and two horizontal sections near the top (H1) and bottom (H2) of the intersection between the support and the concrete area of the beam were analysed (Fig. 6). From the data recorded with Aramis, the rotation of the sections in the concrete area was determined for the different tests and loading stages. The rotations, expressed in radians, are subsequently included in Tables 5 and 6.

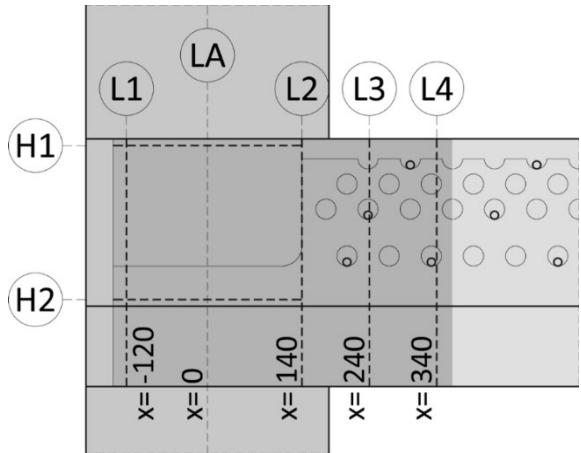

**Fig. 6** *Sections analysed with DIC.*

2.6. Numerical analysis

A direct approach with a 3D FEM model was developed using Ansys Workbench Academic Research V.2018. Numerical analysis aims to evaluate the ability of the model to predict the behaviour obtained in laboratory tests.

The lower glulam flanges were modelled with 3D-volume elements and mesh with the 20-node SOLID186-elements, with three degrees of freedom per node. The glulam was modelled by taking each layer into account, with a rigid coupling applied between layers, disregarding the flexibility of the glue and using the same SOLID186-elements. Steel elements combine SOLID186-elements with SOLID187-elements defined by 10 nodes having three grades of freedom, which are suitable for modelling irregular meshes. The model used CONTA174 to represent contact and sliding between 3D target surfaces (TARGE170), limiting contact restrictions according to the relative behaviour of the bodies.

The concrete was modelled using a Menetrey-Willam [33] strain softening model with the parameters shown in Table 2.

An orthotropic material was considered for timber with a longitudinal elastic modulus $E_L$=11600 MPa according to Eurocode 5 [34]. The relations given by $E_T/E_L$=0.043, $E_R/E_L$=0.078, $G_{LR}/E_L$=0.064, $G_{LT}/E_L$=0.061 and $G_{RT}/E_L$=0.004, where $E_T$ and $E_R$ are the elastic moduli along the radial and the tangential axes and $G_{LR}$, $G_{LT}$ and $G_{RT}$ represent the corresponding shear modulus and Poisson ratio $\mu_{TR}$=0.31, $\mu_{TL}$ 0.02, $\mu_{RL}$=0.03, were considered in accordance with conventional references [35].

Plywood was considered as an orthotropic material with a longitudinal elastic modulus $E_L$=8925 MPa and $E_T$=$E_R$=650 MPa along the radial and the tangential axes. Shear modulus was considered as $G_{LR}$ 1090 MPa and $G_{LT}$=350 MPa and $G_{RT}$=235 MPa, and Poisson´s ratio $\mu_{TR}$=0.382, $\mu_{TL}$=0.20, $\mu_{RL}$=0.22.

A simplified bi-linear behavioural model was used for the stress-strain steel curve, with a constitutive law assumed to be isotropic as well as the material yield criterion. The hardening rule was assumed to be isotropic, the yielding strength $f_y$=500 MPa and the ultimate strength $f_u$=550 MPa, with a Young's modulus of $E_s$=200000 MPa and a Poisson ratio of $\mu_s$=0.3.

*Table 2*
*Concrete properties used in FEM model according to [32]:*

| Propperty | Value | Reference |
|---|---|---|
| Compressive strength | 38.1 MPa | $f_{ck}$ |
| Young´s modulus | 34796 MPa | $E_c$ |
| Mean tensile strength | 3.396 MPa | $f_{ctm}$ |
| Biaxial compressive strength | 44.268 MPa | $f_{2c2}$ |
| Dilatancy Angle | 30º | $\psi$ |
| Strain at peak compression | 0.0024 | $\varepsilon_{cm}$ |
| Plastic strain at peak compression | 0.0013 | $K_{cm}$ |
| Ultimate plastic strain in compression | 0.01 | |
| Relative stress at nonlinear hardening | 0.40 | |
| Residual compressive relative stress | 0.20 | |
| Plastic strain limit in tension | 0.01 | |
| Residual tensile relative stress | 0.20 | |
| Constitutive law according to Menetrey-Willam model | | |

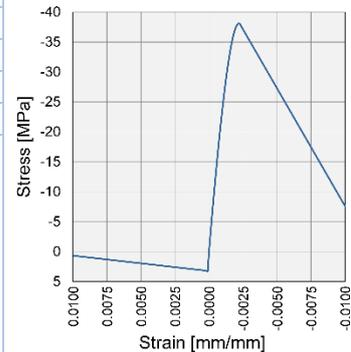

## 3. Results and discussion

### 3.1 Simply supported beam strength

The failure of the beam occurs for a load value of 432.80 kN. During the SS5 test, some cracking was recorded for loads above 180.0 kN, but no external damage was visible in the specimen. During SS6 (failure test), small cracking sounds were heard at loads above 270.0 kN, which continued until the specimen failed. The failure was caused by tensile failure of one of the glulam bottom flanges, which materialised as a brittle failure of the member. As can be seen in Fig. 7, the failure generated a longitudinal crack in a large part of the flange, but in no case was there a failure of adhesion between the glulam flanges and the intermediate plywood boards.

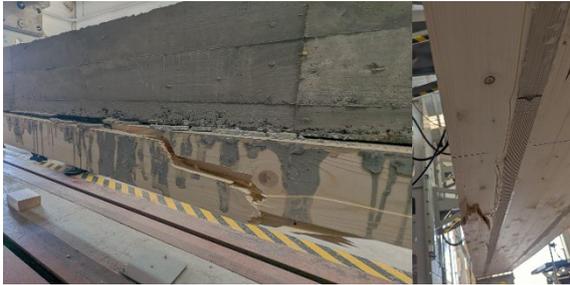

**Fig. 7** *Failure of the simply supported beam in the lower glulam flange.*

Fig. 8 shows the distribution of normal stresses in the section corresponding to the midpoint of the beam. These values were obtained from the finite element model (FEM) by averaging the stress values at the different layers of the cross-section. In order to avoid the stress peaks that occur in the numerical analysis, layers or strata were defined and the average stress values at the different levels were determined for each material. The stresses corresponding to the upper and lower longitudinal reinforcement bars are not plotted in their total value in order not to reduce excessively the scale of representation of the stresses on the timber and concrete, but their value is indicated in the figure for the different load stages considered. At the moment of maximum ultimate load, the average stress in the upper reinforcement bars is 240.24 MPa and 221.17 Mpa in the lower reinforcement bars, which indicates that they remain in their elastic range in all the tests.

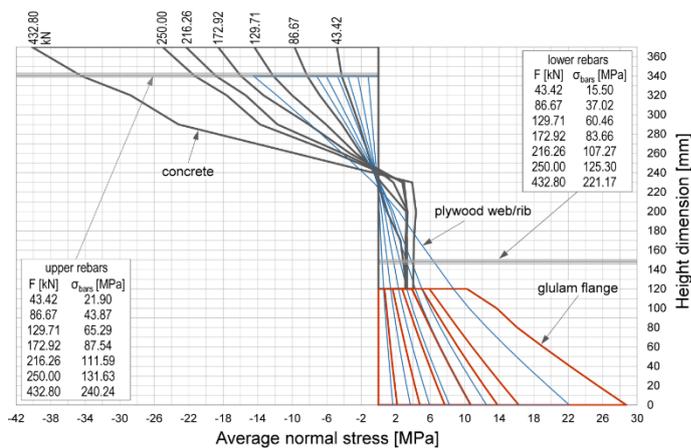

**Fig. 8** *Average stress distribution in the midspan section of the simply supported beam, obtained from the FEM model for different load values. Concrete in black, glulam bottom flange in red and plywood boards in blue.*

According to the FEM model (Fig. 8), only tensile stresses occur in the lower glulam flange, reaching a maximum stress value corresponding to the ultimate load of 28.7 MPa. The plywood board responds to a bending situation, with compressive stresses at the top and tensile stresses at the bottom, reaching maximum values of 14.5 and 22.1 MPa, respectively. These values are lower than the characteristic strength stated by the manufacturer in the case of the boards, but are almost 50% higher than the tensile strength stated for the lower glulam flange. The ultimate load corresponds to a maximum moment of 425.6 kNm, equivalent to that caused by a uniformly distributed load of 97.8 kN/m. Assuming a surface load value between 5.0 and 6.0 kN/m$^2$ for residential or administrative use (which includes around 1.8 kN/m$^2$ of floor self-weight), an overall safety factor between 3.91 and 2.72 would be obtained for floors with spans between 5 and 6 metres.

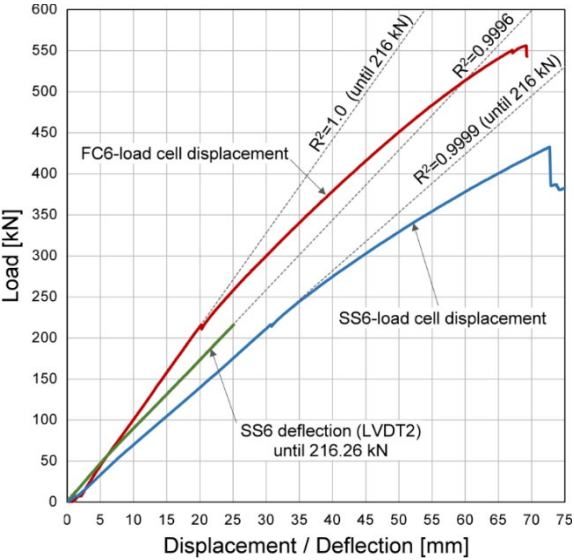

*Fig.9 Deformation (LVDT 2) of the beam in test SS6 and displacement of the load crosshead during test SS6 and test FC6.*

Fig. 9 shows data record corresponding to the LVDT2 sensor during the SS6 test placed at the mid-span point up to a load of 216.26 kN (beam deflection), at which point a stop is made to remove the LVDT with the aim to avoid damage as a consequence of the beam rupture. The linear regression equation corresponding to the LVDT2 record is represented in a dashed line, which shows a practically linear behaviour of the beam up to that point. The curve corresponding to the displacement measured by the load cell, also showed in Fig.9, although it does not allow precise conclusions to be drawn regarding the stiffness of the beam, shows that the linear behaviour is maintained up to a load value of around 250.0 kN. Fig. 8 shows that this load value corresponds to maximum stresses in the concrete around 25.0 MPa, i.e. the beginning of the plasticisation phase of the concrete in accordance with its constitutive law (showed in Table 2).

In relation to the concrete, Fig. 8 shows that the area in contact with the bottom glulam flange reaches tensile values higher than its characteristic strength from the SS1 to the SS6 test. This implies that the concrete is cracked from the first stages of loading. The evolution of these cracks can be seen in Fig. 10, from the images captured with Aramis. The cracks occur from 120 mm of the section, which defines the start of the concrete section above the bottom glulam flange, and extend up to approximately 230 mm, which would correspond to the neutral fibre.

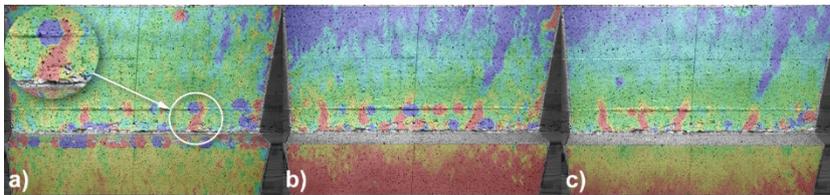

*Fig.10* Aramis images for maximum load in: a) test SS1 (43.28 kN), b) test SS5 (216.40 kN) and c) test SS6 (432.80 kN).

## 3.2 Frame-connected beam strength

The maximum load achieved in the rupture test (FC6) of the frame-connected beam was 556.13 kN. However, the test was stopped without the failure of the specimen being detected, due to significant cracking in the concrete slab anchoring the load cells. Fig. 12 and Fig. 13 show the stress distribution corresponding to the midspan cross-section of the frame-connected beam and to a cross-section 180 mm from the column axis. The stress values were obtained by averaging the results corresponding to the FEM model. It can be seen that the maximum stress values in the lower glulam flange and in the plywood boards corresponding to the webs, 17.64 and 13.67 MPa, respectively, are much lower than those characteristic of the materials used, and significantly lower than those obtained in the midspan section in the case of the simply supported beam, despite the fact that higher load values were achieved for the frame-connected beam. There are several factors that explain this difference. On the one hand, the distribution of moments between the end-joint and the midspan of the beam. Table 3 shows the resulting moments obtained from the stresses shown in Fig. 12 and Fig. 13. It can be seen that the sum of moments at the node and at the centre of the span does not correspond exactly to 100% of the isostatic moment of the simply supported beam. This can be explained by the percentage error assumed in the FEM model, which in this case does not exceed 4%, and by the linear approximation made between the section at 180 mm from the column axis and the theoretical value reached at the axis. On the other hand, it is necessary to consider the lack of continuity of the timber members in the joint area. In the case of the plywood boards, the webs are undercut to allow the passage of the reinforcement that would correspond to a transverse beam in a real situation (Fig. 2). In the case of the lower glulam flange, it is wider than the concrete support, which limits the transmission of compression to the support in this area and concentrates it in the section between the boards (Fig. 11). This way of working of the bottom flange would be more favourable in the case of central multi-span continuous beams.

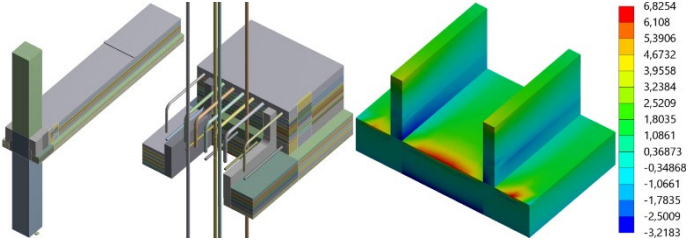

*Fig. 11* *Frame FEM model. Detail of the web plywood boards and the bottom glulam flange in the node area. Stress distribution in a section 180 mm from the support axis at maximum load of FC5.*

From values shown in Table 3, it can be seen how the moment distribution between the node and the mid span varies as the loads increase. The moments were determined from the stress results shown in Fig. 12 and Fig. 13. Initially, the percentage of moment assumed by the node is approximately 43% for a total load of 43.28 kN, but reduces to 35% for a load of 216.40 kN, showing the loss of stiffness of the node. Considering, as indicated for the simply supported beam, that the service load values would not exceed this value, a sizing criterion could be proposed with a moment distribution of one third for the node and two thirds for the centre of the span, respectively. The moments corresponding to test FC6 must be considered with reservations, because only the measurements made with Aramis are available for the FEM model adjustment at this load.

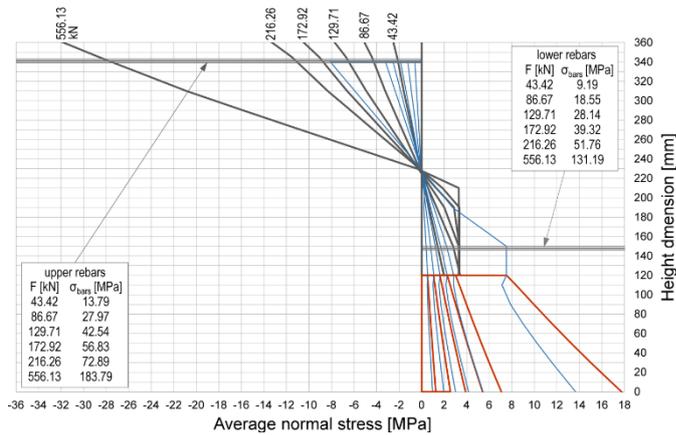

*Fig. 12 Average stress distribution in the cross-section corresponding to the midspan of the frame-connected beam, obtained from the FEM model for different load values. Concrete in black, glulam bottom flange in red and web plywood boards in blue.*

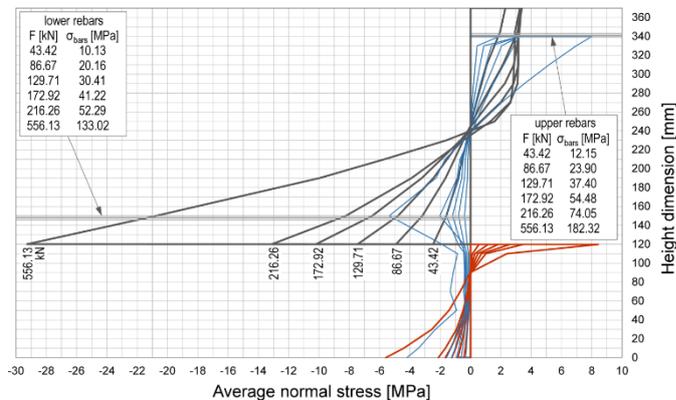

*Fig. 13 Average stress distribution in the frame-connected beam in a cross-section at 180 mm from the column axis, obtained from the FEM model for different load values. Concrete in black, glulam bottom glulam flange in red and web plywood boards in blue.*

Table 3
*Bending moments obtained from the FEM model for the different tests of the portal frame.*

| Test | Maximum load (F) [kN] | Moment corresponding to the simply supported beam [kN·m] | Moment resultant in section at 180 mm from the column axis [kN·m] | Theoretical moment in the axis of the column | | Moment resultant in section at the midspan | |
|---|---|---|---|---|---|---|---|
| | | | | Moment value [kN·m] | Percentage respect to simply supported | Moment value [kN·m] | Percentage respect to simply supported |
| FC1 | 43.28 | 42.56 | 14.53 | 18.44 | 43.3% | 24.26 | 57.0% |
| FC2 | 86.56 | 85.12 | 27.45 | 35.17 | 41.3% | 49.22 | 57.8% |
| FC3 | 129.84 | 127.68 | 37.03 | 48.69 | 38.1% | 78.73 | 61.7% |
| FC4 | 173.12 | 170.23 | 46.02 | 61.28 | 36.0% | 105.43 | 61.9% |
| FC5 | 216.40 | 212.79 | 54.96 | 73.61 | 34.6% | 134.48 | 63.2% |
| FC6 | 556.13 | 546.86 | 122.76 | 166.19 | 30.4% | 308.28 | 56.4% |

Fig. 9 shows the relationship between the load and the displacement of the test crosshead for test FC6, and it is evident that a stop is made when the limit load of the previous test (FC5) is reached. The figure shows a loss of linearity that coincides approximately with this point. In this case, unlike the simply supported beam, this change in behaviour should be related to the level of cracking of the concrete at the top end of the cross-section. This aspect is discussed in more detail in later sections. Despite the high level of cracking reached for the maximum load values, a moment value of around 30% of the isostatic is maintained at the node.

### 3.3 Simply supported beam stiffness

Several tests were carried out to evaluate the in-service behaviour of the specimens, in order to analyse possible stiffness losses related to the behaviour of the concrete and the timber-concrete connection. The different load stages correspond to equivalent uniformly distributed load values from 10 to 50 kN/m which, considering slab spans of 5 to 6 m, would correspond to surface loads between 2.0-10.0 to 1.7-8.3 kN/m$^2$, respectively.

Fig. 14 shows the load-deflection curves, which relate the value of the total applied load to the deflection recorded by the LVDT2 placed at the midspan. The curves show a very linear behaviour of the beam in the successive tests. Only in the case of the SS5 test are there some singularities in the final section, which are discussed below.

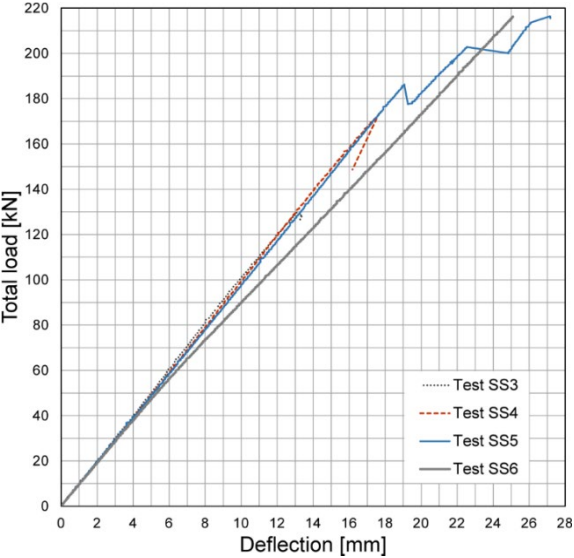

***Fig. 14** Load-deflection curves (LVDT2) for different tests of the simply supported beam.*

Table 4 shows the values of the deflection at the midspan corresponding to the different load stages of the SLS (service limit state) tests performed. From the values recorded by the LVDTs, the equivalent global stiffness (considering the deflection at the mid span) and local stiffness (considering the deflection recorded by the LVDT sensors placed at 0.9 metres from the centre of the span) were determined. In each case, the section of the curve delimited by 10 and 90% of the limit load was considered, with the exception of test SS5 where 80% of the limit load was taken as the upper end, due to the jumps observed in the upper section of the curve. As indicated, SS1 is set as a preload, so it is not considered to calculate the average, which presents a very small deviation for tests SS2 to SS5. The average equivalent stiffness value is 43561 KNm$^2$ for local stiffness and 35852 KNm$^2$ for global stiffness. In addition to bending, the global stiffness is affected by the effect of shear and, therefore, possible relative slip effects in the connection between materials. In this case, the global stiffness is 82% of the local stiffness, which considers only the effect of bending in the central section, indicating a fairly high composite action for the timber-concrete shear connection system.

The overall stiffness corresponding to the data recorded for test SS6 up to a load of 216.40 kN, at which point the central LVDT is removed, was 30483 KNm$^2$. This represents an overall equivalent stiffness loss of 15% compared to the previous tests, and is associated with the jumps seen in the load-deflection curve of test SS5.

*Table 4*
Instantaneous deflection values for different load stages of the simply supported beam and values of the global and local stiffness.

| Test | Maximum load [kN] | Instantaneous deflection [mm] at different load stages | | | | | | | Local stiffness $E \cdot I_{(local)}$ [kN/m$^2$] | Global stiffness $E \cdot I_{(global)}$ [kN/m$^2$] |
|------|------|------|------|------|------|------|------|------|------|------|
| | | 43.28 kN | 86.56 kN | 108.20 kN | 129.84 kN | 155.81 kN | 173.12 kN | 216.40 kN | | |
| **SS2** | 86.56 | 4.17 | 8.78 | - | - | - | - | - | 41981 | 35512 |
| **SS3** | 129.84 | 4.28 | 8.54 | 10.75 | 13.08 | - | - | - | 42081 | 36390 |
| **SS4** | 173.12 | 4.43 | 8.70 | 10.85 | 13.02 | 15.73 | 17.63 | - | 44050 | 35881 |
| **SS5** | 216.40 | 4.45 | 8.88 | 11.07 | 13.27 | 15.89 | 17.66 | 27.16 | 46131 | 35623 |
| | | | | | | | | **Average** | 43561 | 35852 |
| | | | | | | | | **Standard deviation** | 4.5% | 1.1% |

As indicated, both in the FEM model and in the images taken with Aramis, it can be seen that from the SS1 test, the concrete reaches stresses higher than its tensile strength in the area of contact with the lower timber flange. Despite this, the stiffness values recorded do not show any significant changes due to cracking of the concrete. This may be due, on the one hand, to the fact that the cracking is visible from the first test and, on the other hand, to the limited contribution of the tensioned concrete zone to the stiffness of the cross-section due to its proximity to the neutral axis.

In the vibration test carried out prior to the application of external loads on the beam, a stiffness value of 42388 kNm$^2$ and a natural frequency of 16.35 Hz was obtained. This stiffness value corroborates those obtained from local measurements in tests SS2 to SS5, with a difference of less than 3% with respect to the average indicated in Table 4.

In relation to the SS5 test, there is a first jump in the graph (Fig. 13) for a load value of 185.81 kN, with a small loss of load on the beam, and a second jump for a load of 202.79 kN. However, no external damage was observed, so the test continued up to the predicted limit load of 216.40 kN. The analysis of the records corresponding to LVDT sensors 7 and 8, placed at the heads of the beam, showed a singularity coinciding with these load values. The records of these LVDTs were coincident for tests SS1 to SS5 (Fig. 15a and Fig. 15b), with slip values that do not exceed 0.05 mm, which indicates a very rigid behaviour of the timber-concrete shear connection. Coinciding with a load value of 185.81 kN there is a sudden jump in the LVDT 7 record reaching a reading of 0.35 mm, and the same happens for a load value of 202.79 in LVDT 8, reaching a reading of 0.61 mm. Those sudden jumps cannot be associated in any case with a failure of the shear connection between the plywood boards and the concrete, since both the shear failure in the plywood board and the crushing of the concrete cylinders on the board would show a ductile behaviour [14].

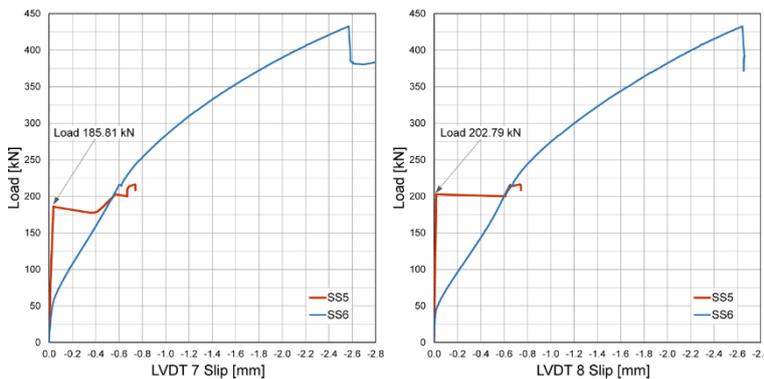

*Fig. 15* Deformation recorded by a) LVDT 7 and b) LVDT8 placed on the heads of the simply supported beam.

In a previous study [14], the shear behaviour of these connections was analysed through push-out tests. It was observed that the friction between the plywood and the concrete provides a non-negligible shear transfer capacity with a very high stiffness. Specifically, specimens with a length of 450 mm, double 21 mm plywood boards without holes and 60 mm depth of concrete slab in contact with the boards were tested. Average failure load values of 42.6 kN and $k_{ser}$=547 kN/mm were obtained. By adding holes to the boards, the failure load varied between 93.4 and 230.6 kN and the $k_{ser}$ between 399 and 171 kN/mm, with an average value of 284 kN/mm. In the load-slip curves corresponding to the connections where holes were drilled in the boards, a first branch with higher stiffness could be seen, which became more ductile from approx. 40 kN onwards. All these factors point to the fact that the jumps seen in the load-strain curve of test SS5 and in the records of LVDT7 and LVDT8 may be due to having reached the limit of the frictional transfer capacity between the timber and the concrete in each of the end boards of the beam, for a load value of 185.81 kN in the first one and for a load value of 202.79 kN in the second one. In the SS6 test, a more ductile behaviour of the shear connection linked to the load transfer through the concrete cylinders that pass through the boards would therefore be appreciated. This raises the possibility of using a primer on the boards in the future to limit the possibility of load transfer to the concrete by friction. In this way, an initial loss of stiffness of around 15% as observed in the SS6 test would have to be assumed, but the possibility of sudden jumps in in-service behaviour such as those recorded in the SS5 test would be eliminated.

Applying the Gamma Method included in EuroCode 5 [34] for the calculation of composite sections, it can be verified that, considering the average $k_{ser}$ value of 284 kN/m, the theoretical equivalent stiffness goes from 47945 kNm$^2$, corresponding to a total composite action, to a value of 34779 kN/m$^2$. This value corresponds to a difference of only 3% with the average global equivalent stiffness value indicated in Table 4. Since, as explained above, the initial global stiffness is somewhat lower than the theoretical one, this would indicate that the $k_{ser}$ must be somewhat higher.

Considering the usual load values for residential and office use [36] between 5 and 6 kN/m$^2$, including self-weight of the floor, permanent and use loads and floor spans from 5 to 6 metres, would result a uniformly distributed load around 25-36 kN/m over the beam, which would be equivalent to point loads with a total value between 108.2-155.81 kN. In other words, in a real service situation we would not exceed the loads applied during the SS4 test and the instantaneous deflection of the element would be between 10.85 and 15.73 mm (between L/460 and L/381).

### 3.4 Frame-connected beam stiffness

Fig. 16a and Fig. 16b show the load-displacement curves for the frame-connected beam, relating the load applied over the beam and the displacement of the midpoint of the span. Contrary to what happened in the case of the simply supported beam, it can be seen that the curves do not have a linear behaviour along their entire length and that the slope of the curves varies throughout the successive tests.

Table 5 shows the values recorded by LVDT 2, corresponding to midpoint for the different tests carried out. For tests FC2 to FC5, the records of the displacement at the column head (LVDT 1) and at the beam end support (LVDT 4) are included, which allows the beam deflection to be determined; also the rotations calculated from Aramis for the different sections defined in section 2.5. are included. Table 6 shows the cross-section deflections for the FC6 test for different load stages.

It was observed that the deflections and rotations corresponding to the FC1 test were higher than those of the subsequent tests. This same effect was observed in the case of the simply supported beam and is related to preload and adjustment between the different materials. Since it occurs for very low load values, it can be considered as a preload that will take place during the construction phase in real structures and has no significant effect on the service behaviour of the system.

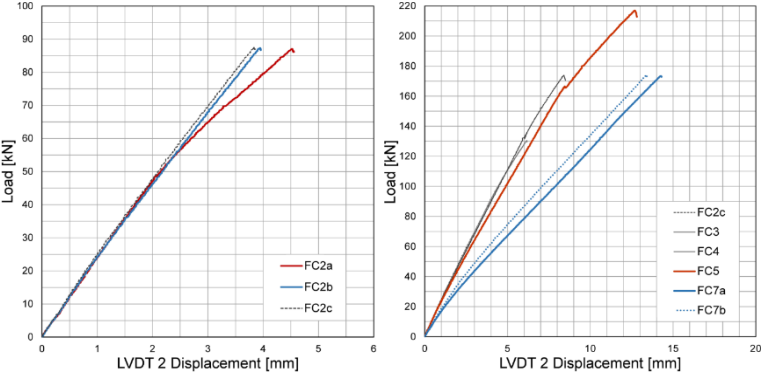

*Fig. 16* Load-displacement curves (LVDT 2) of the frame-connected beam midspan for (a) tests FC2a, FC2b and FC2c and (b) tests FC2c, FC3, FC4, FC5, FC7a and FC7b.

*Table 5*
*LVDT sensors and Aramis measurements for FC2-FC5 and FC7 tests:*

| Test | Data | | Total load | F=43.28 | F=86.56 | F=129.84 | F=173.12 | F=216.40 |
|---|---|---|---|---|---|---|---|---|
| FC2a | LVDT 1 (top support displacement) [mm] | | 0.24 | 0.87 | | | | |
| | LVDT 2 (midpoint displacement) [mm] | | 1.83 | 4.48 | | | | |
| | LVDT 4 (lower face support displacement) [mm] | | 0.19 | 0.48 | | | | |
| | Deflection (difference LVDT2-LDVT4) [mm] | | 1.64 | 4.00 | | | | |
| | Concrete Aramis rotation [rad] | LA rotation | 0.00049 | 0.00126 | | | | |
| | | H1 rotation | 0.00048 | 0.00128 | | | | |
| | | H2 rotation | 0.00048 | 0.00127 | | | | |
| | | L2 rotation (140 mm from the axis) | 0.00056 | 0.00138 | | | | |
| | | L3 rotation (240 mm from the axis) | 0.00061 | 0.00151 | | | | |
| | | L4 rotation (340 mm from the axis) | 0.00066 | 0.00173 | | | | |
| FC2b | LVDT 1 (top support displacement) [mm] | | 0.29 | 0.71 | | | | |
| | LVDT 2 (midpoint displacement) [mm] | | 1.87 | 3.89 | | | | |
| | LVDT 4 (lower face support displacement) [mm] | | 0.21 | 0.43 | | | | |
| | Deflection (difference LVDT2-LDVT4) [mm] | | 1.66 | 3.46 | | | | |
| FC2c | LVDT 1 (top support displacement) [mm] | | 0.29 | 0.72 | | | | |
| | LVDT 2 (midpoint displacement) [mm] | | 1.60 | 3.37 | | | | |
| | LVDT 4 (lower face support displacement) [mm] | | 0.18 | 0.39 | | | | |
| | Deflection (difference LVDT2-LDVT4) [mm] | | 1.62 | 3.39 | | | | |
| | Concrete Aramis rotation [rad] | LA rotation | 0.00049 | 0.00127 | | | | |
| | | H1 rotation | 0.00048 | 0.00128 | | | | |
| | | H2 rotation | 0.00049 | 0.00127 | | | | |
| | | L2 rotation (140 mm from the axis) | 0.00055 | 0.00138 | | | | |
| | | L3 rotation (240 mm from the axis) | 0.00061 | 0.00165 | | | | |
| | | L4 rotation (340 mm from the axis) | 0.00066 | 0.00173 | | | | |
| FC3 | LVDT 1 (top support displacement) [mm] | | 0.31 | 0.72 | 1.09 | | | |
| | LVDT 2 (midpoint displacement) [mm] | | 1.84 | 3.77 | 6.11 | | | |
| | LVDT 4 (lower face support displacement) [mm] | | 0.19 | 0.38 | 0.60 | | | |
| | Deflection (difference LVDT2-LDVT4) [mm] | | 1.65 | 3.39 | 5.51 | | | |
| | Concrete Aramis rotation [rad] | LA rotation | 0.00047 | 0.00099 | 0.00148 | | | |
| | | H1 rotation | 0.00046 | 0.00098 | 0.00150 | | | |
| | | H2 rotation | 0.00049 | 0.00099 | 0.00154 | | | |
| | | L2 rotation (140 mm from the axis) | 0.00052 | 0.00109 | 0.00177 | | | |
| | | L3 rotation (240 mm from the axis) | 0.00062 | 0.00132 | 0.00218 | | | |
| | | L4 rotation (340 mm from the axis) | 0.00065 | 0.00138 | 0.00231 | | | |
| FC4 | LVDT 1 (top support displacement) [mm] | | 0.30 | 0.70 | 1.08 | 1.46 | | |
| | LVDT 2 (midpoint displacement) [mm] | | 1.85 | 3.85 | 5.86 | 8.35 | | |
| | LVDT 4 (lower face support displacement) [mm] | | 0.15 | 0.34 | 0.53 | 0.75 | | |
| | Deflection (difference LVDT2-LDVT4) [mm] | | 1.70 | 3.51 | 5.33 | 7.60 | | |
| | Concrete Aramis rotation [rad] | LA rotation | 0.00045 | 0.00094 | 0.00143 | 0.00197 | | |
| | | H1 rotation | 0.00044 | 0.00093 | 0.00144 | 0.00198 | | |
| | | H2 rotation | 0.00042 | 0.00092 | 0.00143 | 0.00199 | | |
| | | L2 rotation (140 mm from the axis) | 0.00051 | 0.00109 | 0.00169 | 0.00244 | | |
| | | L3 rotation (240 mm from the axis) | 0.00062 | 0.00134 | 0.00208 | 0.00290 | | |
| | | L4 rotation (340 mm from the axis) | 0.00071 | 0.00145 | 0.00225 | 0.00319 | | |
| FC5 | LVDT 1 (top support displacement) [mm] | | 0.32 | 0.72 | 1.11 | 1.47 | 1.80 |
| | LVDT 2 (midpoint displacement) [mm] | | 1.95 | 4.19 | 6.47 | 9.08 | 12.61 |
| | LVDT 4 (lower face support displacement) [mm] | | 0.15 | 0.36 | 0.56 | 0.74 | 0.98 |
| | Deflection (difference LVDT2-LDVT4) [mm] | | 1.80 | 3.83 | 5.91 | 8.34 | 11.63 |
| | Concrete Aramis rotation [rad | LA rotation | 0.00041 | 0.00090 | 0.00141 | 0.00193 | 0.00257 |
| | | H1 rotation | 0.00041 | 0.00090 | 0.00143 | 0.00197 | 0.00264 |
| | | H2 rotation | 0.00043 | 0.00091 | 0.00141 | 0.00195 | 0.00262 |
| | | L2 rotation (140 mm from the axis) | 0.00050 | 0.00111 | 0.00174 | 0.00245 | 0.00346 |
| | | L3 rotation (240 mm from the axis) | 0.00062 | 0.00136 | 0.00213 | 0.00296 | 0.00409 |
| | | L4 rotation (340 mm from the axis) | 0.00068 | 0.00151 | 0.00237 | 0.00332 | 0.00458 |
| FC7a | LVDT 2 (midpoint displacement) [mm] | | 2.60 | 6.00 | 9.62 | 13.31 | |
| FC7b | LVDT 2 (midpoint displacement) [mm] | | 2.99 | 6.64 | 10.45 | 14.20 | |

By comparing the curves in Fig. 16a and the values in Table 5, the influence of the axial load over the supports on the deformation of the beam can be evaluated. The deformation values corresponding to tests FC2b and FC2c, with 100.0 and 200.0 kN of load over each column, respectively, do not show significant differences. This is favourable, as it implies that the balancing effect of the axial load over the supports will occur even for reduced load values on the upper floor levels. On the other hand, although the deflections are clearly higher in test FC2a, when there is no load on the supports, it can be seen that the rotation of the sections is very similar in the three tests. Therefore, it seems that the main difference between the three cases studied is the relative displacement that occurs at the top head of the column, which is significantly reduced in the cases where the supports are loaded. The deflection in the frame-connected beam for a load of 86.56 kN increases by 18% in the case of the test performed without load on the supports (FC2a) in relation to the FC2c test. The effect of the load on the supports is also evident in the vibration measurements carried out as indicated in section 2.4. The measured natural frequency was 26.10 and 28.05 Hz, without and with load on the supports, respectively, which represents in this case a difference of 7%.

*Table 6*
*Aramis results for test FC6.*

| Concrete Aramis rotation [rad] | F= 43.28 | F= 86.56 | F= 129.84 | F= 173.12 | F= 216.40 | F= 259.68 | F= 302.96 | F= 346.24 | F= 389.52 | F= 432.80 | F= 476.08 | F= 519.36 | F= 554.71 |
|---|---|---|---|---|---|---|---|---|---|---|---|---|---|
| LA rotation | 0.00044 | 0.00093 | 0.00145 | 0.00199 | 0.00256 | 0.00318 | 0.00402 | 0.00510 | 0.00641 | 0.00813 | 0.01002 | 0.01246 | 0.01537 |
| H1 rotation | 0.00045 | 0.00095 | 0.00134 | 0.00205 | 0.00265 | 0.00339 | 0.00436 | 0.00555 | 0.00699 | 0.00864 | 0.01053 | 0.01283 | 0.01555 |
| H2 rotation | 0.00043 | 0.00094 | 0.00148 | 0.00204 | 0.00264 | 0.00336 | 0.00432 | 0.00552 | 0.00694 | 0.00859 | 0.01046 | 0.01273 | 0.01543 |
| L2 rotation (140 mm from the axis) | 0.00055 | 0.00121 | 0.00191 | 0.00263 | 0.00343 | 0.00469 | 0.00613 | 0.00774 | 0.00955 | 0.01163 | 0.01402 | 0.01689 | 0.02014 |
| L3 rotation (240 mm from the axis) | 0.00066 | 0.00146 | 0.00229 | 0.00314 | 0.00406 | 0.00543 | 0.00702 | 0.00881 | 0.01087 | 0.01315 | 0.01569 | 0.01865 | 0.02201 |
| L4 rotation (340 mm from the axis) | 0.00801 | 0.00173 | 0.00266 | 0.00360 | 0.00457 | 0.00598 | 0.00774 | 0.00961 | 0.01165 | 0.01407 | 0.01657 | 0.01967 | 0.02318 |

When analysing the results of tests FC2c to FC5 (Fig. 16b), it can be seen how the load-deformation curves have an initially linear section, which tends to curve once the load value of the previous test has been exceeded. This behaviour is related to the successive opening of cracks in the tensioned concrete on the upper face of the beam-to-column node and also entails a slight loss of stiffness in the successive tests, which is evident in the slope of each curve. This aspect has also been highlighted in the analysis of the variation of the moment distribution when increasing load.

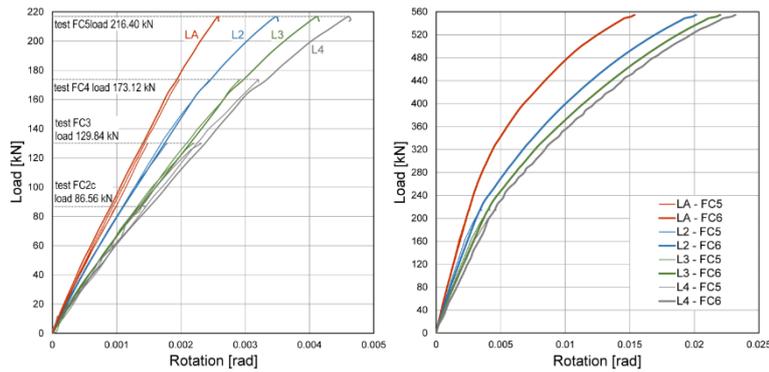

*Fig. 17 Load-rotation curves a) for tests FC2 to FC5 and b) for tests FC5 and FC6.*

The increase in the deformation for the successive tests FC2 to FC5 is reflected in small variations in the rotation of the sections and in the displacement of the head of the top support. From the values in Table 5, it can be seen that this rotation increase does not exceed 5%, and that it is different depending on the section considered. The variation of these rotations can be seen in Fig. 17a, which shows an appreciably linear tracing of the curves, and with not very significant differences between the successive tests. In the case of the section corresponding to the column axis (LA), the load-rotation relationship remains linear throughout the tests, while sections L2 to L4 reflect a slight curvature in the upper section with a pattern similar to that of the load-displacement curves. Fig. 17b shows the comparison between the rotation of the sections in the case of test FC5 and test FC6. In this case, it can be seen how, once again, the load-rotation relationship is practically linear up to the previous load level (216.40 kN) and how the rotation increases from that value onwards. In this case, the loss of linearity in the axis of the support can also be seen. This behaviour can be related to the cracking pattern observed in the images taken with Aramis.

Fig. 18 shows the images taken with Aramis for different load milestones during the FC6 test. Up to a load of approximately 225.0 kN the cracks develop exclusively in the area of the beam adjacent to the support. For a load of approx. 244.0 kN there is clear cracking in the support area, which is reflected in the loss of linearity in the moment-rotation relationship shown in Fig. 16b and with the load-displacement relationship shown in Fig. 9.

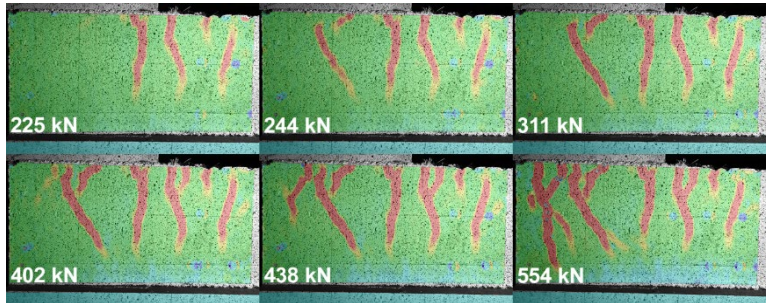

*Fig. 18* Cracking evolution at the portal frame node during the FC6 test.

The results of the high degree of cracking reached during test FC6 are evident in the deformations recorded during tests FC7a and FC7b (Fig. 16b and Table 5). Comparing the results corresponding to tests FC5 and FC7b (both with a load of 200.0 kN on each support), an increase in deformations between 40 and 60.0% is observed. Comparing the results of tests FC7a (with no load on the supports) and FC7b, it is also observed that the effect of the load on the supports is much lower in this case than in the one observed for test FC2. Obviously, this is related to the loss of stiffness in the node, and therefore the lower incidence of the axial force on the rotation of the support in the node.

3.5 Simply supported vs frame-connected.

The beneficial effect of the continuity of the beam connection to the supports is shown by comparing the deflection values recorded for both beams. Fig. 19 shows the LVDT 2 records for tests up to a load value of 216.40 kN. Taking as reference floor spans between 5 and 6 m and total loads between 5.0 and 6.0 kN/m$^2$, the serviceability values of distributed load on the beam would not exceed 36.0 kN/m in any case. This value would correspond to a total point load value of 155.8 kN, considering equivalent deflections for the simply supported beam, which would be below the loading milestone of the SS4 and FC4 tests. On the safety side, the results corresponding to the SS5 and FC5 tests are compared in this section. Additionally, the results corresponding to the SS6 test for the simply supported beam are included, which show the loss of stiffness explained in previous sections; despite the fact that this loss of stiffness, associated with the friction load transmission limit, has occurred for total load values higher than the service value of 155.0 kN.

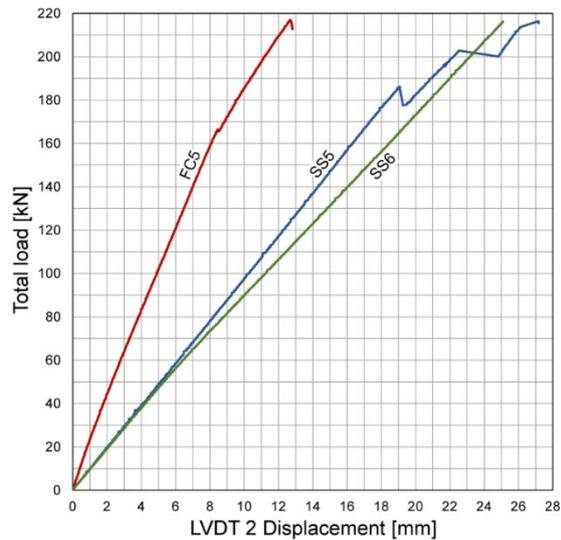

**Fig. 19** *LVDT2 displacements in the simply supported beam and the frame-connected beam.*

Comparing the deflection results of test 5 for the frame-connected beam and the simply supported beam (FC5 and SS5), a reduction in LVDT displacement at the midpoint of the frame beam bigger than 50% is observed. This value is even higher when comparing the deflection value at the frame instead of the mid-point displacement.

A comparison of Fig. 8, Fig. 12 and Fig. 13 also shows a reduction in the level of stresses in the materials, which results in a better use of the materials. Considering the load milestone of 216.25 kN, which would correspond to an overall partial factor of 1.40 compared to the service value of 155.0 kN, the stress level in the timber elements is reduced by 49%.

Vibration measurements showed an increase in natural frequency of 59.6% when comparing the simply supported beam to the frame-connected beam with no load on the supports, rising to a 71.6% increase when the supports are loaded.

In addition, the continuous concrete connection of beams, floors and supports would provide advantages in terms of stiffness against horizontal actions, the diaphragm effect of the floor slab and reduction of shrinkage cracking; aspects that deserve further study.

## 4 Conclusions

The experimental behaviour of a full-scale simply supported beam and a full-scale frame-connected beam with a timber-concrete composite solution in which the shear connection is made by means of perforations in the web plywood boards has been studied. With the experimental results, a numerical model has been validated for the analysis of stresses in the different elements that make up the system. The beams tested had a span of 5.9 metres.

The continuous connection of the beam and the supports reduces the deformation of the frame-connected beam by 50% compared to the simply supported solution. The beam-support continuity joint provides a significant stiffness which reduces the stresses in the timber elements by up to 49%, which are the cause of failure in the case of the simply supported beam. Based on the results obtained, a moment distribution of one third for the end nodes and two thirds for the mid span could be proposed for the portal frame design in the case of service load values corresponding to residential or administrative use.

The continuous connection of the beam to the supports significantly increases the natural frequency of vibration of the beam, up to 71.6% when the supports are loaded, which corresponds to a normal service situation.

The results are promising with regard to the behaviour of the proposed TCC structural system, and should be completed by increasing the number of specimens and by a more detailed study of the design of the node. A field of study is also opened in relation to the behaviour of this system against horizontal actions, the incidence of concrete shrinkage in continuous frames and the effect of the creep deformation of the materials.

Acknowledgments/Funding

This study is part of the research project of I+D+i PID2022-138649OB-I00 "Comprehensive industrialized system of high energy efficiency for frame structures of timber-recycled concrete (FrameTCC), funded by MCIN/AEI/10.13039/501100011033/ and "FEDER A way of making Europe".